# Paradoxical Quantum Scattering off a Time Dependent Potential?

Ori Reinhardt and Moshe Schwartz

*Department of Physics, Raymond and Beverly Sackler Faculty Exact Sciences,
Tel Aviv University, Tel Aviv 69978, Israel*

We consider the quantum scattering off a time dependent barrier in one dimension. Our initial state is a right going eigenstate of the Hamiltonian at time t=0. It consists of a plane wave incoming from the left, a reflected plane wave on the left of the barrier and a transmitted wave on its right. We find that at later times, the evolving wave function has a finite overlap with left going eigenstates of the Hamiltonian at time t=0. For simplicity we present an exact result for a time dependent delta function potential. Then we show that our result is not an artifact of that specific choice of the potential. This surprising result does not agree with our interpretation of the eigenstates of the Hamiltonian at time t=0. A numerical study of evolving wave packets, does not find any corresponding real effect. Namely, we do not see on the right hand side of the barrier any evidence for a left going packet. Our conclusion is thus that the intriguing result mentioned above is intriguing only due to the semantics of the interpretation.

The quantum problem of an incoming plain wave encountering a static potential barrier in one dimension is an old canonical text book example in which the reflection and the transmission of the particle are defined and calculated [1-10]. The problem of a time dependent barrier has been explored within time dependent perturbation theory [6-15] and more recently some exact results appeared in the literature [16,17] making clear that new interesting results may be expected when time dependent potentials are involved [18]. The purpose of the present manuscript is to present a new intriguing result in that field. We show that starting with a state which is traditionally interpreted as a particle travelling from left to right, partly reflected and partly transmitted through the barrier, we end up with a finite probability of finding the system in a state of the opposite nature. Namely, there is a finite probability to find a state which is interpreted as a particle travelling from right to left, partly reflected back to the right and partly transmitted through the barrier.

Consider first the quantum problem of a particle in the presence of a time independent barrier (see Fig. 1). The Hamiltonian of the system is given by

$$H = -\frac{\hbar^2}{2m}\frac{\partial^2}{\partial x^2} + V(x), \qquad (1)$$

where $V(x) = 0$ for $|x| > a$. An example for such a potential is depicted in Fig. 1.

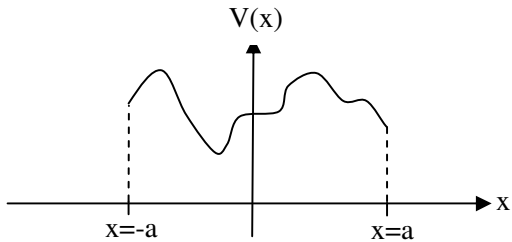

Fig. 1. A general potential $V(x)$ that vanishes for $|x| > a$.

The eigenstates of the Hamiltonian (Eq. 1) can be classified according to the absolute value of incoming momentum and to whether they are right or left going. We define the right going eigenstates for $k > 0$:

$$\Psi_k^+(x) = \begin{cases} e^{ikx} + A_k^+ e^{-ikx} & \text{for } x < -a \\ D_k^+ e^{ikx} & \text{for } x > a \end{cases}, \qquad (2)$$

where the solution within the segment $|x| \leq a$ is not relevant to our discussion, although it affects the coefficients $A$ and $D$. The first term in the region $x < -a$ is interpreted as the wave incoming from the left, the second term is interpreted as the reflected wave and the single term in the region $x > a$ is interpreted as the transmitted wave. Similar left going eigenstates are:

$$\Psi_k^-(x) = \begin{cases} e^{-ikx} + A_k^- e^{ikx} & \text{for } x > a \\ D_k^- e^{-ikx} & \text{for } x < -a \end{cases}. \qquad (3)$$

For a symmetric potential $D^+ = D^- = D$ and $A^+ = A^- = A$. In the following paragraphs we prefer to work with a delta function potential, $V_0 \delta(x)$, because of the simplicity of the analysis, though, as we will show later the intriguing result we obtain is generic. In that case

$$A_k = \frac{\beta}{-\beta + ik} \quad \text{and} \quad D_k = \frac{ik}{-\beta + ik}, \qquad (4)$$

where $\beta = \frac{mV_0}{\hbar^2}$. The transmission is given by

$$T = \frac{k^2}{\beta^2 + k^2} \qquad (5)$$

Now we consider a time dependent delta function potential, $\lambda(t)\delta(x)$. The set $\{\Psi_k^\sigma(x) \mid k > 0, \sigma = \pm\}$, which are eigenstates of the Hamiltonian at time $t = 0$ is complete. The scalar products of any two states in the basis is given in bra-ket notation by

$$\langle \Psi_k^\sigma | \Psi_l^\tau \rangle = 2\pi \delta(k - l) \delta_{\sigma\tau}, \qquad (6)$$

thus the state at a general time can be written as

$$|\Psi(t)\rangle = \int_0^\infty dl \left[ \exp\left(-\frac{iE_l t}{\hbar}\right) \right] \sum_{\tau=+}^{-} c_l^\tau(t) |\Psi_l^\tau\rangle. \quad (7)$$

The equation for the coefficients is readily obtained from the time dependent Schrödinger equation.

$$\frac{dc_k^\sigma}{dt} = \frac{1}{2\pi i\hbar} D_k^*[\lambda(t) - \lambda(0)] \int_0^\infty dl D_l \exp\left(\frac{i(E_k - E_l)t}{\hbar}\right) \sum_{\tau=+}^{-} c_l^\tau(t) \quad (8)$$

It is thus clear that $\frac{dc_k^\sigma}{dt}$ does not depend on $\sigma$. Consider next the initial condition

$$|\Psi(0)\rangle = |\Psi_q^+\rangle. \quad (9)$$

Since the potential changes in time it is expected that the particle will exchange energy with the potential and thus at later times the expansion of the potential in terms of the original eigenstates will contain contributions with $l \neq q$. The intuitive feeling based on the interpretation of the state $|\Psi_q^+\rangle$ as a wave coming from the left partly reflected back and partly transmitted through the barrier, suggests, however, that we should have zero overlap at any time with states having negative $\sigma$, which correspond following the same interpretation to waves coming from the right. The exact result that $\frac{dc_k^\sigma}{dt}$ does not depend on $\sigma$ proves the opposite:

$$c_k^-(t) = c_k^+(t) - \delta(k-q). \quad (10)$$

To show that this result is not an artifact of our choice of the delta potential, we repeat the same calculation for a time dependent potential, $V(x,t)$, which vanishes for $|x| > a$.

The equations for the coefficients are given in the general case by

$$\frac{dc_k^\sigma}{dt} = \frac{1}{2\pi i\hbar} \int_0^\infty dl \left[ \exp\left(\frac{i(E_k - E_l)t}{\hbar}\right) \right] \sum_{\tau=+}^{-} \langle \Psi_k^\sigma | \Delta V | \Psi_l^\tau \rangle c_l^\tau(t), \quad (11)$$

where $\Delta V = V(x,t) - V(x,0)$.

For the $\delta$ function, the matrix element $\langle \Psi_k^\sigma | \Delta V | \Psi_l^\tau \rangle$ does not depend on $\sigma$ (and on $\tau$) and that is the reason that $\frac{dc_k^\sigma}{dt}$ does not depend on $\sigma$. Now the situation is different. To get oriented consider $\frac{dc_k^\sigma}{dt}$ at very short times for the initial state $|\Psi_q^+\rangle$. We will approximate $\Delta V = \frac{\partial V}{\partial t}(x,0)t$. For very short times $t$ and for $k$'s such that $\frac{|E_k - E_q|t}{\hbar} \ll 1$,

$$\frac{dc_k^\sigma}{tdt} \approx \frac{1}{2\pi i\hbar} \left\langle \Psi_k^\sigma \left| \frac{\partial V}{\partial t}(x,0) \right| \Psi_q^+ \right\rangle. \quad (12)$$

Since the matrix elements $\left\langle \Psi_k^\sigma \left| \frac{\partial V}{\partial t}(x,0) \right| \Psi_l^\tau \right\rangle$ do not vanish identically for $\sigma \neq \tau$,

$$c_k^-(t) = \frac{1}{4\pi i\hbar} \left\langle \Psi_k^- \left| \frac{\partial V}{\partial t}(x,0) \right| \Psi_q^+ \right\rangle t^2 \neq 0. \quad (13)$$

Once $c_k^-(0)$ does not vanish at short times it will not do so at later times. It is interesting to compare $c_k^-(t)$ with $c_k^+(t)$ for $k \neq q$ and very short time $t$. To keep matters simple let us think about a smooth symmetric potential and $E_q$ smaller than the typical height of the barrier. The wave functions $\Psi_q^+(x)$ and $\Psi_k^+(x)$ decay within the barrier from left to right with rates, $\gamma_q$ and $\gamma_k$ respectively and the wave functions $\Psi_q^-(x)$ and $\Psi_k^-(x)$ decay from right to left with the same rates, $\gamma_q$ and $\gamma_k$. Therefore, we may expect that the very short time ratio $\left|\frac{c_k^-}{c_k^+}\right|$ decays as a function of the width of the potential as

$$\left|\frac{c_k^-(t)}{c_k^+(t)}\right| \propto \frac{2\sinh\left[(\gamma_k - \gamma_q)a\right]}{(\gamma_k - \gamma_q)(\gamma_k + \gamma_q)} \exp\left[-(\gamma_k + \gamma_q)a\right], \quad (14)$$

when for both $\gamma$'s, $\gamma a \gg 1$. Thus as the barrier tends to become macroscopic the ratio above tends exponentially to zero and the strange waves incoming from the right disappear relative to the expected ones as a response to the time dependence of the potential. Can we say something about the time dependence of that ratio? Equation (11) can be taken as a starting point for a time dependent perturbation expansion. It is clear that for a potential of the form, say,

$$V(x,t) = \lambda(t)V(x), \quad (15)$$

and for $k \neq q$, the ratio $\left|\frac{c_k^-(t)}{c_k^+(t)}\right|$ does not depend on time, within first order time dependent perturbation theory. This implies that although there is a finite transition rate into the strange waves incoming from the right it is exponentially small in the width of the barrier.

The result presented here implies that it is necessary to consider wave packets in order to verify or disprove the existence of a real effect. Therefore we solve numerically the Schrödinger equation for the time dependent delta potential barrier. This solution gives the time dependent solution for every initial wave function. For simplification, we suppress in the following, the parameters $m$ and $\hbar$ in the Schrödinger equation by inserting a general diffusion coefficient $\eta$. We consider an initial state which is a normalized Gaussian of the form

$$\psi(x,t=0) = \left(\frac{1}{2\pi\sigma_0^2}\right)^{\frac{1}{4}} e^{-\frac{(x-x_0)^2}{4\sigma_0^2}} e^{ik_0 x}. \quad (16)$$

For $x_0 < 0$ and $k_0 > 0$ this is a wave packet starting on the left and travelling from left to right. Consider first the evolution of the wave packet above in the presence of a time independent delta barrier. The probability density and probability current of the evolving wave function are given in Fig. 2.

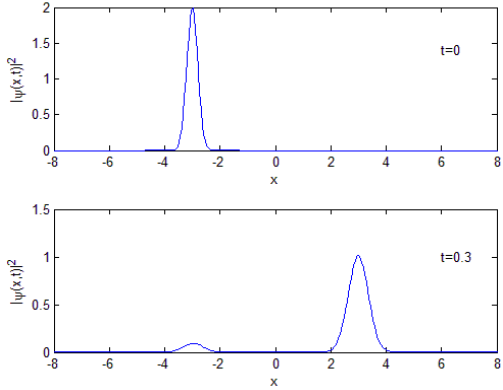

Fig. 2. The evolution of a wave packet encountering a delta barrier. The initial state is a normalized Gaussian with $x_0 = -3, \sigma_0 = 0.2, k_0 = 50$. The delta barrier has an amplitude of $\lambda = 6$. The diffusion coefficient is $\eta = 0.2$.

It can be seen that similar to the traditional interpretation, where we talked about the reflected plane wave and the transmitted one, we obtain here also after the arrival of the initial packet at the barrier a transmitted packet moving to the right and a reflected packet moving to the left, which look very much Gaussian. It is interesting to note that the total probability to find the particle on the right of the barrier, as obtained by numerical integration, is equal, indeed, to the transmission calculated for plane waves, with $k_0$ replacing $k$.

Now the interesting question is what will be the effect of a time dependent potential? Is it possible to see any earmark of the seemingly paradoxical result we have mentioned previously? An example for an evolving wave function in the presence of a time dependent potential is shown in Fig 3, where the time dependent part of the potential was taken as

$$\lambda(t) = \lambda\left(1 + \alpha \sin(\omega_0 t)\right). \quad (17)$$

We can see from Fig. 3. that the arriving, reflected and transmitted wave packets are seen very clearly, though they are not pure Gaussians any more but have undergone some modulation. Most importantly, it can be seen that there is no evidence of anything travelling from right to left on the right hand side of the barrier. This was studied rather carefully running many solutions, finding no real effect of the intriguing result mentioned earlier.

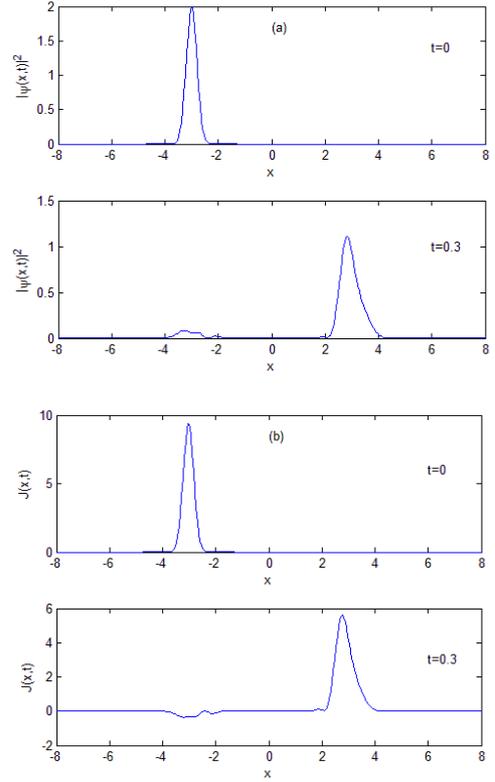

Fig. 3. The evolution of a wave function encountering an oscillating delta barrier. (a) Probability density $|\psi(x,t)|^2$ (b) Probability current $J(x,t)$. The initial state is a Gaussian wave packet with $x_0 = -3, \sigma_0 = 0.2, k_0 = 50$. The parameters of the time dependent barrier (Eq. 17) are $\lambda = 5, \alpha = 1, \omega_0 = 7 \cdot (2\pi/0.3)$. The diffusion coefficient is $\eta = 0.2$.

To check on the numerical procedure and to obtain an analytic expression for the evolving wave packet, it is also interesting to look at this problem using first order perturbation theory. An analytical expression has been derived from equation (8) for the first order $c_k^\pm$ coefficients in the case of an initial state of the form given by equation (16) with a time dependent potential given by equation (17),

$$c_k^{\pm(1)}(t) = \frac{\lambda\alpha\sigma_0\sqrt{\pi}}{(2\pi)^2}\left(\frac{1}{2\pi\sigma_0^2}\right)^{\frac{1}{4}} e^{ik_0 x_0}\frac{ik}{\beta+ik}\int_0^\infty dl \frac{l^2}{\beta^2+l^2} \cdot$$

$$\cdot\left[e^{-\sigma_0^2(k_0-l)^2-ilx_0}\left(\beta\cdot erf\left(-\frac{x_0}{2\sigma_0}-i\sigma_0(k_0-l)\right)+il\right)+\right.$$

$$\left.+e^{-\sigma_0^2(k_0+l)^2+ilx_0}\left(-\beta\cdot erf\left(-\frac{x_0}{2\sigma_0}-i\sigma_0(k_0+l)\right)+il\right)\right]\cdot$$

$$\cdot\left[\frac{e^{i(\omega_0+\eta(k^2-l^2))t}-1}{\omega_0+\eta(k^2-l^2)}+\frac{e^{-i(\omega_0-\eta(k^2-l^2))t}-1}{\omega_0-\eta(k^2-l^2)}\right]$$

(18)

The first order correction to the time dependent wave function due to the oscillating part is

$$\psi^{(1)}(x,t) = \int_0^\infty dk \cdot e^{-i\eta k^2 t} c_k^{\pm(1)}(t)\left(e^{-ik|x|}+e^{2i\tan^{-1}\left(\frac{k}{\beta}\right)}e^{ik|x|}\right).$$

(19)

In the following graph we compare the analytic first order correction to the wave function with the corresponding numerical result for a specific time. We consider a relatively small coupling and we see good agreement.

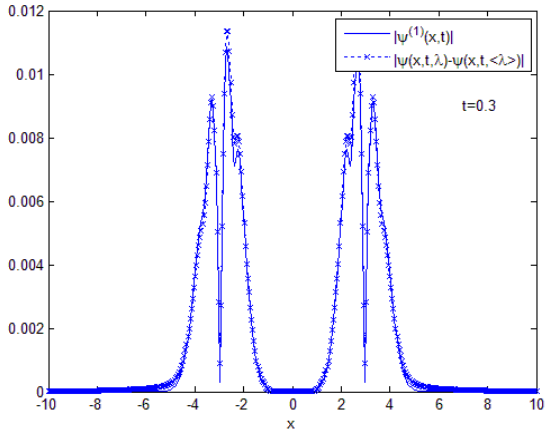

Fig. 4. The first order perturbation theory wave function. Continuous line: analytical expression, dashed line: numerical solution expression. The initial state is a Gaussian wave packet with $x_0=-3, \sigma_0=0.2, k_0=50$. The delta barrier has the parameters $\lambda=3, \alpha=0.1, \omega_0=7\cdot(2\pi/0.3)$. The diffusion coefficient is $\eta=0.2$.

To conclude, we have shown that a time dependent barrier on which particles are scattered from the left generates particles that seem to arrive from the right and being scattered off the barrier. By numerically investigating the time dependent Schrödinger equation with the initial state being a wave packet travelling from left to right, we found no evidence of wave packets coming from the other side, as seemingly implied by the former mentioned analytical result. This was shown numerically for a delta potential and joins our previous statement that the effect disappears when the potential is wide (14).

This seeming contradiction most probably relates to the fact that the traditional interpretation of the non-localized states as a sum of the original incoming wave, a transmitted and a reflected wave, is still classical.